%% file: Kineti_E.tex
\documentclass{rcdj}

\newcommand{\mes}{{\rm mes}\,}

\begin{document}
\input rcdhyph

\setcounter{page}{235}

\author{V.\,V.\,KOZLOV}
\address{V.\,V.\,KOZLOV}
{Department of Mechanics and Mathematics\\
Moscow State University,
Vorob'ievy Gory\\
119899, Moscow, Russia\\
E-mail: vako@mech.math.msu.su}
\title{KINETICS OF COLLISIONLESS CONTINUOUS MEDIUM}
\smalltitle{KINETICS OF COLLISIONLESS CONTINUOUS MEDIUM}
\journal{REGULAR AND CHAOTIC DYNAMICS, V.\,6, \No3, 2001}
\acce{In this article we develop Poincar\'e ideas about a heat
balance of ideal
gas considered as a collisionless continuous medium. We obtain the
theorems on diffusion in nondegenerate completely integrable systems. As a
corollary we show that for any initial distribution the gas will be
eventually irreversibly and uniformly distributed over all volume,
although every particle during this process approaches arbitrarily close
to the initial position indefinitely many times. However, such individual
returnability is not uniform, which results in diffusion in a reversible
and conservative system. Balancing of pressure and internal energy of
ideal gas is proved, the formulas for limit values of these quantities are
given and the classical law for ideal gas in a heat balance is deduced. It
is shown that the increase of entropy of gas under the adiabatic extension
follows from the law of motion of a collisionless continuous medium.}
\amsmsc{37H10, 70F45}
\doi{10.1070/RD2001v006n03ABEH000175}
\date{June 8, 2001}

\maketitle

\section{Heat balance}

The establishment of heat balance of gas in a vessel is one of the central
problems of nonequilibrium statistical mechanics. The conventional model
is Boltzmann\f Gibbs gas: ensemble of a large (but finite) number of
identical solid balls elastically colliding with each other and with the
walls of the vessel. According to the classical approach based on the
Boltzmann kinetic equation the process starts with practically
instantaneous establishment of the Maxwell velocity distribution, and then
(not so fast and with oscillations) the gas density becomes
balanced~\cite{1}.

Unfortunately, such approach involve several fundamental difficulties.
First, the Boltzmann equation is approximate. It does not take into
account multiple collisions, and, besides this approach assumes the
statistical independence of the number of pair collisions. This assumption
(Stosszahlansatz by P.\,Ehrenfest and T.\,Ehrenfest~\cite{2}) is
plausible, but it certainly does not follow immediately from the dynamics
of the Boltzmann\f Gibbs gas model. Furthermore, there are difficulties in
an adjustment of the solutions of the Boltzmann equation with the
reversibility property of the dynamics equations and with the Poincar\'e
theorem on returning (see~\cite{3,4} for the discussion of these
problems).

Logical opportunity of the adjustment of the irreversible behavior of a
system with the properties of reversibility and returnability was shown by
M.\,Kac~\cite{4,5} on a so-called {\it circular model}, which, however, is
not related to the gas theory. A phase space in the Kac model is an
ensemble of white and black balls in vertexes of regular $n$\1gon.
Besides, a set~$M$ of vertexes of $n$\1gon consisting of $m<n/2$ elements
is selected. The dynamics of the circular model is determined by rotation
on one element counter-clockwise. If a ball does not belong to the
set~$M$, its color does not change, and if it belongs to~$M$, the color of
the ball changes to the opposite. The dynamics of such system is clearly
invariant relatively to the direction of rotation (reversibility). It is
also possible to verify that after~$2n$ rotations the system will turn
into the initial state (returnability).

Let $N_c(t)$ and $N_b(t)$ be the numbers of black and white balls in
integer moments of time~$t$. As Kac has proved, the mean value of the
ratio
$$
  \Bigl\langle \frac {N_c (t) -N_b (t)} {n} \Bigr\rangle,
$$
calculated over all possible states of the set~$M$, decreases for
$n\to\infty$ as $(1-2\mu)^t$, where $\mu$ is a limit value of ratio~$m/n$.
Thus, if $\mu<1/2$ then after the large enough period of time the average
number of white and the average number of black balls will coincide.

The Kac circular model is an advanced version of earlier Ehrenfests'
model~\cite{2}, that possessed only the property of reversibility. The
model also shows a characteristic feature of the conventional of reasoning
in statistical mechanics: evaluation of average values, passage to the
limit on the number of particles $(n\to\infty)$, and then passage to the
limit on time $(t\to\infty)$. The latter limit is connected with the fact
that the average returning time  tends to infinity together with~$n$,
hence one should choose a time interval, which is less than the order of
this quantity.

Actually, the justification of thermodynamics involves additional
difficulties of a different kind. The matter is that {\it the ideal gas}
is considered as a system of noninteracting particles. In particular, they
cannot collide with each other. It is under this assumption that the
perfect gas law is deduced in statistical mechanics. On the contrary, when
the interaction is taken into account (in particular, assuming the
possibility of collisions is ), then using the canonical Gibbs
distribution we obtain the equation of state, different from the classical
Clapeyron equation (see~\cite{6}).

On the other hand, as it is shown in papers~\cite{7,8}, the Clapeyron
equation is deduced from the general principles of statistical mechanics
under the assumption that the density of distribution of probabilities is
a single-valued function of the total energy of system of particles. It
should be emphasized that this function does not necessarily coincide with
the density of Maxwell distribution.

The ideal gas is the fundamental model of mechanics of a continuous medium
and statistical mechanics. Therefore, the problem of justification of
irreversible behavior of ideal gas that does not require the Boltzmann
mechanism of pair collisions gains a special importance. And it is not at
all obvious that such irreversibility mechanism actually exists.

These problems constitute the subject of the present paper.

\section{The ideal gas as a collisionless continuous medium}

The above mentioned difficulties can be overcome if the ideal gas is
considered as {\it a collisionless continuous medium}. It is necessary to
note that the hypothesis on the continuity property of gas is in good
agreement with the continuity of velocity distribution of gas particles.

Besides, collisionless models play an essential role in many parts of
mathematical physics. As an example, the theory by Ya.\,B.\,Zeldovich
could be mentioned that explains an occurrence of inhomogeneities in a
distribution of pulverulent substance in the Universe (see~\cite{9}).
Another important example is the Burgers equation. It describes the
dynamics of fluid without pressure and is one of possible simplifications
of the Navier\f Stokes equations~\cite{10}. Multidimensional hydrodynamics
of invariant manifolds of Hamiltonian systems, developed in work~\cite
{11}, also describes the evolution of a collisionless medium.

For the first time, the ideal gas has, apparently, been considered as a
collisionless continuous medium by H.\,Poincar\'{e} in~\cite{12}. He
studied the behavior of ideal gas in a rectangular parallelepiped
$$
  \Pi^n = \{0\le z_1\le l_1\dts 0\le z_n\le l_n \}
$$
($l_s$ being the edges of the parallelepiped) for $n=1,\,2$ and 3.
Poincar\'{e} called such gas {\it one-dimensional}. In Poincar\'{e}'s
terminology, {\it three-dimensional gas} is consisted of molecules, which
can collide with each other. His basic observation was that, independently
of the initial distribution, gas eventually tends to uniform filling
of~$\Pi$. Thus, the ideal gas shows the irreversible behavior. Every
particle of gas approaches arbitrarily close to the initial position
infinitely many times. However, because of nonuniformity of the
returnability property, a nonreversible {\it diffusion} of gas occurs.
Besides, the equations of motion of a collisionless medium are invariant
under reflection of time $t\mapsto -t $. Thus, as far back as in 1906
Poincare showed on the simplified model (directly related to the kinetic
theory) the compatibility of the reversibility and returnability
properties with irreversible behavior of a dynamical system.

Unfortunately, these remarkable ideas of Poincar\'{e} were not properly
understood and remained unclaimed. I have not found any work on
statistical mechanics that mentions his ideas in connection with the
problem of irreversibility. The comments to the Poincar\'{e}'s work of
1906 in volume III of the Russian edition of his collected works
completely miss the point.

Some interesting works on the kinetic theory, that also use the model of
collisionless medium, have appeared recently. As an example we shall
mention {\it the dynamical demon of Maxwell}~\cite{13,14}. However, they
do not refer to the Poincar\'{e}'s pioneer works.

Of course, Poincar\'{e} has considered only the most simple variants and
his works do not contain precise statements with complete and rigorous
proofs in modern understanding of these words. However, his ideas finally
(after almost 100 years) deserve an involvement into the area of study of
modern nonequilibrium statistical mechanics. The purpose of the present
paper is the development of the Poincar\'{e}'s ideas on heat balance of
ideal gas as a collisionless continuous medium.

So, we consider the dynamics of particles in a $n$-dimensional
parallelepiped~$\Pi^n $. Clearly,~$\Pi^n$ allows a natural $2^n$\1leaf
covering by $n$\1dimensional torus $\mT^n = \{x_1\dts x_n \mod 2\pi \} $
with branching on the boundary of~$\Pi^n $. Variables~$x$ and~$z$ are
related as follows: $x =\pi z/l$ if~$z$ increases from~0 up to~$l$, and
$x=2\pi-\pi z/l$ if~$z$ decreases from~$l$ to~0 (Fig.~1).

\fig<bb=0 0 91.2mm 53.0mm>{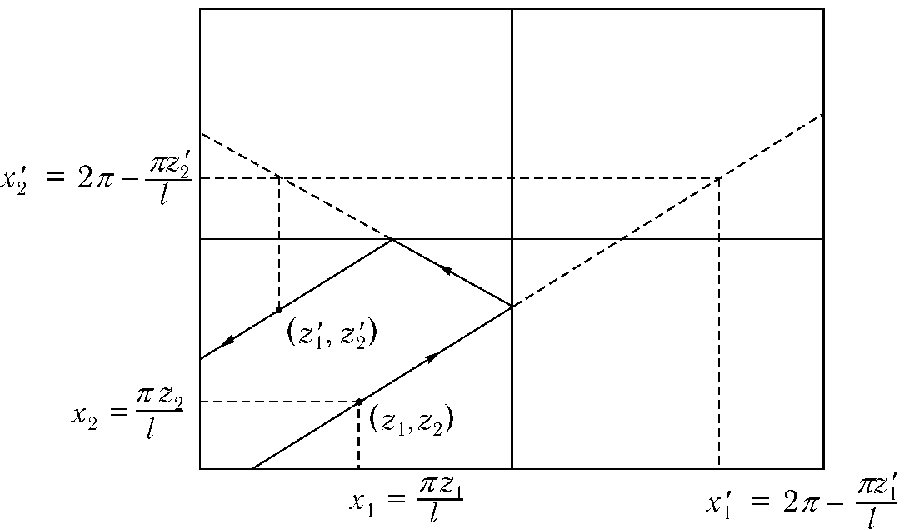}

Let $v_1\dts v_n$ be the velocity components of a gas particle in
$\Pi^n\subset\mR^n=\{z\}$. Then the rates of variation of its
$x$\1coordinate are equal to
$$
  \om_1 =\frac {\pi v_1} {l_1} \dts \om_n =\frac {\pi v_n} {l_n}.
  \eqno {(2.1)}
$$
Hence, in variables $x\mod 2\pi,\,\om$, the dynamics of gas particles
in~$\Pi$ is described by the equations
$$
  \dot x_s =\om_s, \q \dot\om_s=0\q (s=1\dts n).
  \eqno {(2.2)}
$$

\section{The first theorem on diffusion}

Equations (2.2) describe an evolution of the integrable system. A phase
space
$$
  \mP^{2n} = \mP^n\times \mT^n = \{\om, \, x\mod 2\pi \}
$$
is foliated on invariant tori $\om = (\om_1\dts \om_n) = \const $, which
are filled with conditionally periodic trajectories with frequencies
$\om_1\dts\om_n$. For almost all $\om\in\mR^n$ these trajectories are
everywhere dense (and, actually, uniformly distributed) on torus $\om
=\const$.

Such picture is generally characteristic for completely integrable
Hamiltonian systems with compact energy surfaces~\cite{15}. In a
neighborhood of invariant tori it is possible to introduce the
action-angle variables $x\mod 2\pi,\, y$; in these variables the
Hamiltonian equations take the form:
$$
  \dot y_s=0, \q \dot x_s =\om_s (y); \q 1\le s\le n.
  \eqno{(3.1)}
$$
In a {\it nondegenerate} case, when
$$
  \pt {(\om_1\dts \om_n)} {(y_1\dts y_n)} \ne 0,
$$
it is possible to change from variables $y$ to new variables~$\om$.
In these coordinates equations~(3.1) have form~(2.2).
Thus, equations~(2.2) represent {\it a universal} form of equations of motion of nondegenerate completely integrable systems.

Let $f (\om,\, x)$ be a Lebesgue integrable function, which is $2\pi
$\1periodic on each of coordinates $x_1\dts x_n$. Let $g\colon \mT^n\to\mR
$ be a Riemann integrable function; in particular, it means that it is
restricted. Let's introduce the following function of time
$$
K(t) = \intl_{\mR^n} \intl_{\mT^n} f(\om,\,x-\om t) g(x)\,d^nx\,d^n\om.
\eqno {(3.2)}
$$
Since the function $f(\om,\,x-\om t)$ is Lebesgue integrable on all values
of~$t$, and $g$ is measurable and restricted function, integral~(3.2) is
correctly defined.

The function $K(t)$ has an obvious interpretation. First of all we note
that it follows from equations~(2.2) that $x-\om t=x_0 =\const$. Let $f\ge
0$ be (according to Gibbs) the density of distribution of integrable
systems in~$\mP$ (density of probability measure), and $g$ be the
characteristic function of Jordan measurable domain~$D$ on~$\mT^n$. It is
clear that
$$
  \langle f \rangle = \intl_{\mP^n} f\, d^nx\, d^n\om=1.
$$
It is easy to see, that in this case~$K(t)$ is equal to a fraction of all
the systems whith phases (i.\,e. the $x$\1coordinates of points
on~$\mT^n$) belonging to the domain~$D$.

Let's study the behavior of function $K(t)$ for $t\to\om\infty $.

\begin{teo}\label{teo1}
There exists
$$
  \lim_{t\to\pm\infty} K(t) = \langle f \rangle \ol{g},
  \eqno {(3.3)}
$$
where
$$
  \ol{(\cdot)} = \frac{1}{(2\pi)^n} \intl_{\mT^n}(\cdot)\, d^nx.
$$
\end{teo}

Let's return to the case, where $f$ is the density of distribution of
probability measure, and~$g$ is the characteristic function of measurable
domain~$D$. Then relation~(3.3) transforms into the equality
$$
\lim_{t\to\pm\infty} K(t) = \frac{\mes G}{\mes \mT^n}.
\eqno {(3.4)}
$$
Hence, independently of the initial distribution after an unbounded period
of time the system becomes evenly distributed on phases. This result shows
the irreversible diffusion in nondegenerate integrable systems.

We divide the proof of theorem~\ref{teo1} onto several items. Since a
Lebesgue integrable function can be presented as a difference of two
non-negative integrable functions, we shall assume $f\ge 0$.

1) Let $g(x)=\ol{g}=\const $. Then
$$
  \intl_{\mP} f(\om,\, x-\om t) g\, d^n x\, d^n\om =\langle f \rangle \ol{g}
$$
using the formula of change of variables in a multiple integral.

2) Let $g(x)=\exp i(m,\, x)$, $m\in\mZ^n$, and $m\ne 0$. By setting
$u=x-\om t $, we obtain
$$
  K(t) = \intl_{\mR^n} f_{-m} (\om) e^{i (m, \om) t} d^n\om,
$$
where
$$
  f_{-m} = \intl_{\mT^n} f(\om, \, u) e^{i (m, u)} d^n u
$$
is a Fourier coefficient of function $f$, considered as a function
on~$\mT^n$, multiplied by~$(2\pi)^n$. The Fubini theorem implies that the
function $f_{-m} \colon \mR^n\to \mR$ is Lebesgue integrable. Hence
(according to the theory of Fourier transform), $K(t)\to 0$ for
$t\to\infty$.

3) Items 1 and 2 imply that theorem~\ref{teo1} is valid for any
trigonometric polynomial~$g$.

4) We shall now use the well-known statement from the theory of Riemann
integral (compare with~\cite{16}). Let function $g\colon \mT^n\to \mR $ be
Riemann integrable. Then, for all $\eps>0$, there exist two trigonometric
polynomials~$g_1$ and~$g_2$ such that
$$
  \alig{
  \t{(a)} & \q g_1 (x) \le g (x) \le g_2 (x) \; \t {for all} \; x\in \mT^n, \\
  \t{(b)} & \q\ol {g}_2-\ol{g} _1 < \eps.
 }
$$
The proof of this statement uses the Weierstrass approximation theorem.

5) Let
$$
  K_j(t)=\intl_{\mP} f(\om,\,x-\om t) g_j(x) d^nx\,d^\om;\q j=1, \, 2.
$$
Since $f\ge 0$, then for all $t$,
$$
  K_1(t) \le K(t) \le K_2(t).
$$
According to item~3, for $t\to\infty $ the difference $K_2(t)-K_1(t)$
tends to
$$
  \langle f \rangle (\ol{g}_2-\ol{g}_1)<\langle f \rangle\eps.
  \eqno {(3.5)}
$$
Here we use property (b) from item~4.

Thus, inequality (3.5) holds for all $t>T_1(\eps)$.

6) According to item~3, for all $\eps>0$ there exists $T_2(\eps)$ such that for all $t>T_2(\eps)$ we have
$$
|K_j(t) -\langle f \rangle\ol {g}_j|<\eps.
\eqno {(3.6)}
$$
On the other hand, using property (a) and inequality $f\ge0$ we obtain
$$
\langle f \rangle\ol {g}_1\le \langle f \rangle\ol {g} \le
\langle f\rangle\ol {g}_2.
\eqno {(3.7)}
$$
Inequalities (3.5)--(3.7) imply that for $t>\max(T_1(\eps),\, T_2(\eps))$ the inequality is fulfilled
$$
  |K(t) -\langle f \rangle\ol{g}|< 3\langle f \rangle\eps
\qed
$$

These arguments are similar to the proof of the Weyl uniform distribution
theorem~\cite{16}. As a matter of fact, Poincar\'{e} does not give the
precise formulation of the statement on the limit uniform phase
distribution. He considers only item~2 in the special case of $n=1$,
assuming that the function~$f$ is continuously differentiable with respect
to~$\om$.

\section{Equalization of density}

Let $f(\om, \, x) \ge 0 $ be the density of distribution of gas particles
in a phase space for $t=0$, and~$g(z)$ be the characteristic function of a
measurable domain~$G$ in~$\Pi^n$. By definition of density, $\langle
f\rangle=1$. On the other hand, using the explicit formulas for $2^n
$\1leaf covering $\mT^n\to\Pi^n$, we obtain
$$
  \ol{g} = \frac{1}{l_1\ldots l_n} \intl_{\Pi^n} g (z)\, d^nz=
  \frac{\mes G}{\mes\Pi}.
  \eqno {(4.1)}
$$

Hence, by theorem~\ref{teo1},
$$
  \lim_{t\to\pm\infty} K(t) = \frac{\mes G}{\mes\Pi}.
$$

Thus, after some period of time $t$ a fraction of gas particles, which are
situated in the domain~$ {G\subset\Pi}$, is proportional to the volume
of~$G$. So, we come to the following result (formulated by
Poincar\'{e}~\cite{12}): independently of the initial distribution of~$f$,
for ${t\to +\infty}$ and ${t\to-\infty}$ the density of gas in the
vessel~$\Pi$ will be irreversibly equalized. Let's emphasize once again
that the diffusion of a collisionless medium is determined by
nonuniformity of returnability of its particles to the initial positions.

Since the difference $K(t)-\mes G/\mes\Pi$ tends to zero, fluctuations of
density of ideal gas decrease unrestrictedly with a course of time. For
the Boltzmann\f Gibbs gas consisting of {\it finite} number of particles, it
is necessary, at first, to give rigorous definition of density
equalization. Unfortunately, no precise theoretical results were presented
for this subject so far. However, in any case (because of the Poincar\'{e}
theorem on returning), equalization of densities of the Boltzmann\f Gibbs
gas should be accompanied by sustained fluctuations. Problems of numerical
simulation of Boltzmann\f Gibbs gas are discussed, for example, in
paper~\cite{17}.

To estimate the rate of equalization of density of ideal gas, we shall
consider a simple example. We assume that the velocities of gas particles
are subjected to the normal distribution:
$$
   f(\om,\,x)=\frac{\lm(x)}{(\sqrt{2\pi}\sg)^n}e^{-\frac{\om^2}{2\sg^2}},
  \eqno {(4.2)}
$$
$$
  \om^2 =\om_1^2 +\ldots +\om_n^2.
$$
Here $ \lm $ is the non-negative measurable function on~$\mT^n$, and
$$
\intl _ {\mT^n} \lm (x) \, d^nx=1.
\eqno {(4.3)}
$$
If we interpret (4.2) as the density of Maxwell distribution then the
variance~$\sg^2$ is in proportion to Kelvin temperature~$\tau$.

In the case in question we have the following equality:
$$
  K(t) -\frac {\mes G} {\mes \Pi} = {\sum_m}'g_m\lm_{-m}\intl_{-\infty}^{\infty}
   \frac{e^{-\frac{\om^2}{2\sg^2}}}{(\sqrt{2\pi}\sg)^n} e^{i (m, \om) t} d^n\om,
  \eqno {(4.4)}
$$
where $g_m$ is a Fourier coefficient of the lifting of function~$g(z)$
on~$\mT^n$, and $(2\pi)^n\lm_m$ is a Fourier coefficient of function
$\lm\colon\mT^n\to\mR$. By formula~(4.3), we obtain a simple estimate
$|\lm_m|\le 1$. On the other hand, using formula~(4.1) we have the
following inequality:
$$
|g_m |\le \frac {\mes G} {\mes \Pi}.
$$

Hence, from (4.4) we obtain the inequality
$$
  \Bigl|K(t)-\frac{\mes G}{\mes \Pi}\Bigr |\le\frac{\mes G}{\mes \Pi}
  \sum_{m\ne 0} e^{-\frac{\sg^2 t^2 (m, m)}{2}}.
  \eqno {(4.5)}
$$
The series on the right hand side converges for all $t>0$ (if $\sg\ne 0$)
and its sum tends to zero extremely fast, when $t\to\pm\infty$.

The sum of a majorizing series
$$
    \sum_{m\ne 0} e^{-\frac {\sg^2 t^2 (m, m)}{2}}
    \eqno {(4.6)}
$$
is expressed in theta-functions. It is equal to
$$
  [\ta_3, \, (0, \, q)] ^n-1,
$$
where $q =\exp(-\sg^2t^2/2)$, and the third theta-function is defined by
the series
$$
  \ta_3 (v, \, q) = \sum_{-\infty}^\infty q^{n^2} e^{i2\pi nv}.
$$
It is clear, that $\ta_3(v,\,q)\to 1$ for $q\to 0$.

Estimate (4.5) is universal in a sense that it does not contain the
function~$\lm$. In particular, it is possible to use Dirac's
$\dl$\1function as~$\lm$. In that case, the gas at the initial moment is
concentrated in one point (by the way, this situation does not contradict
the hypothesis about collisionlessness of the medium). Since~$\sg^2$ is
proportional to~$\tau$, series~(4.6) actually depends on the
combination~$t\sqrt{\tau}$. Therefore, the duration of density
equalization process decreases with the growth of temperature
as~$1/\sqrt{\tau}$.

Note that in contrast to Kac's model (and to the conventional knowledge
about the mechanism of heat balance in gases) the equalization of density
occurs without preliminary averaging with respect to the states and
without determination.

\section{The second theorem on diffusion}

Let functions $f,\, g\colon \mR^n\times \mT^n\to \mR $ are integrable
together with their squares (with belong to class~$L_2(\mP)$). It is clear
that for all values of~$t$ the function $f(\om,\, x-\om t)$ also belongs
to~$L_2$. Therefore, the function
$$
  K(t) = \intl_{\mP} f(\om, \, x-\om t) g(\om, \, x) \, d^nx \, d^n\om.
$$
is correctly defined.

\begin{teo}\label{teo2}
Under the above assumptions we have
$$
  \lim_{t\to\pm\infty} K(t) = (2\pi)^n \intl_{\mR^n} \ol{f} \ol{g} \, d^n\om.
  \eqno{(5.1)}
$$
\end{teo}

This result, certainly, does not follow from theorem~\ref{teo1} (as well
as theorem~\ref{teo1} is not a corollary of theorem~\ref{teo2}). In
Poincar\'{e}'s paper~\cite{12} formula~(5.1) is not mentioned.

Before we prove theorem~\ref{teo2}, we shall make one auxiliary statement.
Let
$$
\sum f_m (\om) e^{i (m, x)} \q\t {and} \q \sum g_m (\om) e^{i (m, x)}
\eqno {(5.2)}
$$
be Fourier series of functions $f$ and $g$ for a fixed value of~$\om$. The
Fubini theorem imply that these series are defined for almost all
$\om\in\mR^n$. Moreover, for almost all~$\om$ the functions~$f$ and~$g$
belong to~$L_2(\mT^n)$. Hence,
$$
   \intl_{\mT^n} f(\om, \, x-\om t) g(\om, \, x) \, d^nx
   = (2\pi)^n\sum_m f_m g_{-m} e^{-i (m, \om) t}.
   \eqno {(5.3)}
$$

Since $f,\, g\in L_2$, then the functions $|f_m g_{-m}|$ are integrable
in~$\mR^n$ for all $m\in\mZ^n$. Let's denote $g_m' =g_m\exp[i (m, \, \om)
t]$. It is clear that $g_m'g_{-m}'=g_m g_{-m}$.

\begin{lem} \label{lem1}
$$
\sum_m\intl_{\mR^n}|f_mg_{-m}'+f_mg_m'|\,d^n\om<
\intl_{\mP} (f^2+g^2)\,d^nx\,d^n\om.
\eqno{(5.4)}
$$
\end{lem}

It is a variant of the Bessel inequality. The inequality shows, in
particular, that the series in the left-hand part of~(5.4) converges
uniformly on $t\in\mR$.

\proof*(of the lemma.) We shall use an obvious inequality
$$
  |f_m g_{-m}' +f_{-m} g_m'| \le f_m f_{-m} +g_m g_{-m}.
$$
Since $f_m$ and $f_{-m}$ are complex conjugate, $f_m f_{-m}\ge 0$.
Similarly, $g_m g_{-m}\ge 0$.

Now we are to prove the inequality
$$
\sum_m \intl_{\mR^n} f_m f_{-m} \, d^n\om\le \intl_{\mP} f^2 \, d^nx \, d^n\om.
\eqno{(5.5)}
$$
The similar inequality holds for function $g$.

Indeed, let $f_N$ be the finite sum of the terms of Fourier series~(5.2),
such that $|m|<N$. Then
$$
   0\le\intl_{\mP}(f-f_N)^2\,d^nx\,d^n\om=\intl_{\mP}f^2\,d^nx\,d^n\om-
  \sum_{| m | < N} \intl_{\mP^n} f_m f_{-m} \, d^n\om.
$$
Hence, inequality (5.5) is valid for any finite sum of the series in the
left-hand part of~(5.3). We obtain the required statement by passage to
the limit for $|m|\to\infty$.

Now, let's prove theorem~\ref{teo2}. By lemma~\ref{lem1} (using the
well-known theorems by Levy and Lebesgue)  series~(5.3) converges for
almost all~$\om$, and it is possible to integrate the series term-by-term.
Integrating both parts of equality~(5.3) over~$\mR^n$, we obtain the
relation
$$
  K(t) = (2\pi)^n \intl_{\mR^n} \ol{f} \ol{g} d^n\om
  + (2\pi)^n\sum_{m\ne 0} \intl_{\mR^n} f_m(\om) g_{-m} (\om) e^{-i(m,\om)t}
  d^n\om.
$$

Since functions $f_m g_{-m}$ are Lebesgue integrable, each term of series
in the right-hand side tends to zero, when $t\to\pm\infty $. According to
the lemma, this series converges uniformly on~$t$. Therefore for any
$\eps>0$ there exists~$N(\eps)$, such that the sum of terms of series with
the indices $|m|>N(\eps)$ is less than~$\eps/2$ for all values of~$t$. The
finite sum of remaining terms tends to zero when $t\to\pm\infty$. Hence,
there exists~$T(\eps)$ (actually depending on~$N(\eps)$), such that for
$|t|>T(\eps)$ this sum will be less than~$\eps/2$. So, for $|t|> T(\eps)$
the sum of series is less than~$\eps$, q.e.d.

Theorem~\ref{teo2} make it possible to solve the problem of evolution of
density of distribution of~$f$ for $t\to\infty$. At first, the density
$f(\om,\, x-\om t)$ seems to oscillate conditionally-periodically and,
therefore, there is no limit for $t\to\infty$. However, the density of
distribution of probabilities does not "exist" by itself, but only as an
averaging of some fixed function from~$L_2$. Therefore, the evolution
of~$f$ for $t\to\infty$ should be considered in {\it the generalized}
sense, as it is usually done in the theory of the generalized functions
(see, for example,~\cite{18}).

To understand, how integrable system (2.2) is distributed in phase
space~$\mP$ for $t\to\infty$, we shall introduce the characteristic
function~$g$ of the following set
$$
  G = \{x\mod 2\pi, \, \om\colon x_s'\le x_s\le x_s", \,\om_s'\le \om_s\le \om_s", \, 1\le s\le n \}.
$$
It is clear that
$$
  \langle g \rangle =\left \{\arr [cl]{
(2\pi)^{-n} \ds\prod_{s=1}^n (x_s'-x_s), & \t{if}\; \om'\le\om\le\om", \\
0 & \t{for the remaining} \; \om.
 } \right.
$$

It is asserted that for $t\to\infty$ function $f(\om,\, x-\om t)$
converges weakly to~$\ol{f}$. Limit density~$\ol{f}$ is an integrable
non-negative function on~$\mR^n$, and $\langle \ol{f} \rangle=1 $.

Indeed, by theorem~\ref{teo2}, for $t\to\infty$
$$
  \arr {
  \intl_{\mP} f(\om, \, x-\om t) g(\om, \, x) \, d^nx \, d^n\om\to (2\pi)^n
  \intl_{\mR^n} \ol{f} \ol{g}\, d^n\om = \\
  = \prod_{s=1}^n (x_s''-x_s')\intl_{\om_1'}^{\om_1''}\ldots
   \intl_{\om_n'}^{\om_n''}\ol{f}\,d\om_1\ldots d\om_n.
 }
$$
But exactly the same result is obtained by direct computation of the
average value of limiting density~$\ol{f}$ over domain~$G$.

As an example, we consider {\it a nondegenerate} Hamiltonian system with
one degree of freedom and show that any function of distribution
from~$L_2$ converges weakly to the function depending only on the total
energy. Let's recall the definition of nondegeneracy. We assume that the
whole phase space consists of finite number of pieces, invariant
relatively to the phase flow; on each of such pieces it is possible to
introduce action-angle variables~$y,\,x \mod 2\pi$. The transition from
usual canonical variables~$p,\,q$ to variables~$x,\,y$ is a symplectic
transformation: its Jacobian is equal to one. In new variables,
Hamiltonian~$H(p,\,q)$ depends only on~$y$. We call the system
nondegenerate if
$$
\frac {d^2H} {dy^2} \ne 0
\eqno {(5.6)}
$$
on each of the invariant pieces. It is possible to verify, for example,
that a usual pendulum in the gravity field satisfies these requirements.

In the action-angle variables the Hamiltonian equations
$$
  \dot p =-\pt {H} {q}, \q \dot q =\pt {H} {p}
$$
become
$$
  \dot x =\om (y), \q \dot y=0,
  \eqno {(5.7)}
$$
where $\om =\frac {dH} {dy}$ is the frequency of periodic motion.

According to (5.6), $ \frac {d\om} {dy} \ne 0 $. Hence, $\om $ is a
monotonic function of~$y$ and, consequently, it is possible to change from
variable~$y$ to frequency~$\om$ on each invariant piece. Then,
equation~(5.7) takes the universal form
$$
  \dot x =\om, \qq \dot\om=0
$$
and we can use theorem~\ref{teo2}: the initial density~$f$ tends to the
"equilibrium" limit~$\ol{f}$, depending only on~$\om$. By frequency, the
limit density is a function of~$y$ and, hence, of~$H$.

The attempts to prove that for $t\to\infty $ the distribution of
probabilities tends (in some sense) to the stationary state that
corresponds to the heat balance~\cite{19} (chapter~XII) can be traced as
far back as to Gibbs. According to M.\,Kac~\cite{4} (chapter~III), the
idea that probability should be introduced into mechanics {\it only} by
means of the initial density, certainly seems to be very attractive. But,
generally speaking, this point of view is, apparently, unfounded and the
probability should also appear in mechanics by other ways.

In our opinion, theorems~\ref{teo1} and~\ref{teo2} on diffusion in
integrable systems show the fruitfulness of Gibbs' approach and indicate
that his ideas have not yet been fully realized.

\section{Pressure, internal energy and the equation of state}

Theorem~\ref{teo1} establishes the law of equalization of density of ideal
gas in a rectangular parallelepiped. Theorem~\ref{teo2} let us prove the
equalization of pressure and density of energy of ideal gas, specify the
formulas for the limit values of these quantities and, thus, deduce the
equation of state of gas in heat balance.

First, let's deduce the formula for the pressure of gas on one of the
walls from collisions of the gas particles with this wall. The deduction
of the formula for the pressure follows the classical reasoning in the
elementary kinetic theory of gas (see, for example,~\cite{20}; it is
usually assumed that the gas is already in heat balance and the velocity
distribution  is the Maxwell distribution).

For the sake of clarity, consider a wall, determined by equation
$z_1=l_1$. Let's choose an infinitely small rectangle~$\sg$ with the
center in point ($l_1,\, z_2,\, z_3$), its area being equal to
$d\sg=dz_2\,dz_3$. The particles of gas that can hit rectangle~$\sg $ with
velocity ${\om = (\om_1,\,\om_2,\,\om_3)}$ in the moment of time~$t$ were
situated in the moment of time ${t-dt}$ on the parallel rectangle~$\sg$
with the same area~$\d\sg$ (Fig.~2), during period~$dt$ they sweep volume
$$
  dv =\om_1 \, dt \, dz_2 \, dz_3.
$$

\wfig<bb=0 0 39.9mm 51.2mm>{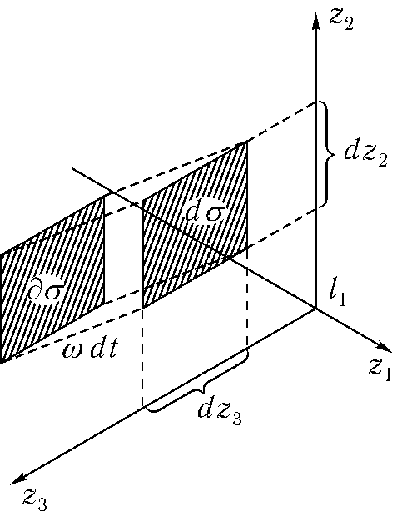}

The number of such particles is equal to
$$
  dn=Nf (\om, \, l_1, \, z_2, \, z_3) \, dv,
$$
where the constant coefficient $N$ is equal to ``the number of particles of
gas in the vessel''. Let $m$ be the mass of ``one'' particle. Then $mN$ is
the total mass of gas. Numbers~$m$ and~$N$ are of the conventional
character, they are introduced for the purpose of comparison of the
formulas obtained below with the known formulas of statistical mechanics.
Since $ \langle f\rangle=1 $, then $mNf$ is the density distribution of
the mass of gas.

During time $dt$ the particles deliver the following impulse to the wall
$$
  dP=2m\om_1 \, dn=2mN\om_1^2f \, dt \, d\sg.
$$
It is clear that $dP/dt$ is a force of pressure. If we divide it by the
area~$d\sg$, then we shall obtain the elementary pressure on the wall in
the point with coordinates $l_1, \, z_2, \, z_3 $. Integrating over all
velocities, we shall obtain the total pressure
$$
  p=2mN\intl_0^\infty \intl_{-\infty}^\infty\intl_{-\infty}^\infty
  \om_1^2 f(\om,\,l_1,\,z_2,\,z_3)\,d\om_1\,d\om_2\,d\om_3.
  \eqno {(6.1)}
$$

Let $f$ be an even function of velocities: $f(-\om,\, z)=f(\om,\, z)$.
Then formula~(6.1) becomes more symmetric:
$$
p=mN\intl_{\mR^3} \om_1^2f (\om, \, l_1, \, z_2, \, z_3) \, d^3\om.
\eqno {(6.2)}
$$

Let's emphasize, that in the initial moment of time the pressure $p$ depends on a point on a wall $z_1=l_1 $.
The formula for pressure on the other walls is the same as~(6.2), with only change of~$\om_1^2$ with~$\om_2^2 $ and~$\om_3^2 $ accordingly.

\begin{rem}\label{rem1}
How shall we determine the pressure of gas in an arbitrary interior point
$z\in\Pi$ on small surface with a normal vector~$n$? If particles of gas
hit the surface from the opposite sides, then (in accordance with the
assumption of evenness of the distribution function) we shall, obviously,
obtain zero pressure. If we put a body of small volume into the
vessel~$\Pi$, then the particles of gas will hit only one side of the
surface. Hence, at this moment of time the pressure can be determined by
the slightly improved formula~(6.2):
$$
  p=mN\intl_{\mR^3} (\om,\,n)^2 f(\om, \, z)\,d^3\om.
$$

Now let $t $ go to infinity. Then, according to section 5, density~$f$
tends to ``the equilibrium'' state~$\ol{f}$, depending only on~$\om$. As a
result, the pressure does not depend on the point of the wall any more:
$$
  p=mN\intl_{\mR^3} \om_1^2 \ol {f} \, d^3\om.
  \eqno {(6.3)}
$$
\end{rem}

\begin{rem}\label{rem2}
Actually formula (6.3) requires additional justification, since in~(6.2)
there is no averaging by configurational space, and function
$\om\mapsto\om_1^2$,  certainly, does not belong to class~$L_2$. We should
understand formula~(6.3) differently.

Let $f(\om,\, x)$ be a density function, "lifted" on direct product
$\mR^3\times\mT^3$. Let's assume
$$
  p(t)=mN\intl_{\mR^3}\om_1^2 f(\om,\, x-\om t)\, d^3\om.
$$
Further, let
$$
  \sum_k f_k(\om) e^{i (k, x)}
$$
be a Fourier series of function $f$, defined for almost all $\om\in \mR^3$.
Then
$$
  p=mN\intl_{\mR^3}\om_1^2\ol {f}\, d^3\om
  +mN\sum_{k\ne 0}e^{i (k, x)}\intl_{\mR^3}\om_1^2 f_k (\om) e^{-i(k,\om)t}
  d^3\om.
$$
If the functions $\om_1^2 f_k$ are summable for all $k\in\mZ^3
$,then~$p(t)$ tends to~(6.3) for $t\to\infty$.
\end{rem}

Note, if $\ol{f}$ depends on $\om^2 =\sum \om_s^2 $, then {\it the
Pascal's law} is valid: the pressure in all directions is identical. In
the opposite case it is not so: a fraction of the gas particles  moving in
different directions is not the same.

For the average kinetic energy of one particle we have the formula
$$
  e = \intl_{\mR^3} \intl_\Pi \frac {m\om^2}{2} f(\om,\, z)\, d^3z\, d^3\om.
$$
If we change to $\sg$\1leaf covering $\mT^3\to\Pi$, we can write this
formula in the following form:
$$
   e=\frac{l_1l_2l_3}{(2\pi)^3}\intl_{\mR^3}\intl_{\mT^3}\frac{m\om^2}{2}
  f (\om, \, x) \, d^3x \, d^3\om.
$$

The average (interior) energy of the whole gas is, obviously, equal to
$$
 E=\frac{mNv}{2}\intl_{\mR^3}\om^2\ol{f}\,d^3\om,
\eqno {(6.4)}
$$
where $v$ is the volume of the vessel. The integral on the right-hand side
converges if the values of pressure of gas are finite on each wall of the
vessel.

If $\ol{f}$ depends on $\om^2$, then (6.3) and (6.4) imply the simple
relation
$$
E = \frac 32 pv.
\eqno {(6.5)}
$$
This formula is well known in the theory of ideal gas. Let
$$
  \Om=dE+p \, dv
  \eqno {(6.6)}
$$
be a 1\1form of heat inflow; it is not a total differential. In accordance
with {\it the law of degradation of energy} of thermodynamics, for~$\Om$
there exists an integrating multiplier: $\bt\Om=dS$, where $S$ is an
entropy, $\bt=1/(k\tau)$, $\tau$ is Kelvin temperature, $k$ is the
Boltzmann constant. According to~(6.5) the integrating multiplier for
form~(6.6) is equal to $(pv)^{-1}$. Hence, $pv $ is proportional to Kelvin
temperature~$\tau$, and we obtain the perfect gas law (the Clapeyron
equation).

It is possible to obtain this result differently, assuming
$$
   \ol{f}=c\rho\Bigl(\bt\frac{m\om^2}{2}\Bigr),
  \eqno {(6.7)}
$$
where $\bt=1/(k\tau)$, $k$ is the Boltzmann constant, $c$ is a normalized
multiplier:
$$
   c^{-1}=v\intl_{\mR^3}\rho\Bigl(\bt\frac{m\om^2}{2}\Bigr)\,d^3\om.
  \eqno {(6.8)}
$$
The distributions of form (6.7) were studied in paper~\cite{8}; the
classical Maxwell distribution also belongs to this type.

After a change $\om_s =\wt\om_s/\sqrt {m\bt}$ we obtain
$$
   c^{-1}=\frac{v}{(\sqrt{m\bt})^3}\intl_{\mR^3}\rho\Bigl(\frac{\wt\om^2}{2}\Bigr)
  \, d^3\wt\om,
$$
$$
  E = \varkappa\frac {Nk\tau} {2}, \q \varkappa =\frac {\intl_{\mR^3} \wt\om^2\rho
   \Bigl(\frac{\wt\om^2}{2}\Bigr)\,d^3\wt\om}{\intl_{\mR^3}\rho
   \Bigl(\frac{\wt\om^2}{2}\Bigr)\,d^3\wt\om}.
$$
For the Maxwell distribution $\varkappa=1$. In paper~\cite{8} a class of
non-Maxwell distributions was described, for which the equality $
\varkappa=1 $ also holds.

\section{Entropy}

As it is known, the entropy is determined by equality
$$
  S = -\intl_{\mP} f(\om,\, x)\ln f(\om, \, x)\, d^nx\, d^n\om,
  \eqno {(7.1)}
$$
where $f$ is a function of distribution of probabilities. For the Maxwell
distribution it coincides with the notion of entropy used in the
equilibrium thermodynamics.

The question is, what is the evolution in time of the entropy in the case
in question, when the gas is represented as a collisionless continuous
medium. It is important to emphasize, that the evolution of state of gas
is {\it an adiabatic process}: there is no energy transfer.

To express $S$ as a function of time, it is necessary to substitute~$x$
with $x-\om t$ in the integrand of~(7.1). However, for such substitution
integral~(7.1) does not change and~$S$ as a function of time is constant.

This simple observation corresponds to the Poincar\'{e}'s result~\cite{12}
that the {\it fine} entropy of mathematicians, in contrast to the {\it
rough} entropy of physicists, is always constant. By the way, the division
of the entropy into the fine and the rough corresponds essentially to the
fine-grain structure and coarse-grain structure of the phase space, that
was introduced by~T.\,Ehrenfest and~P.\,Ehrenfest in their well-known
work~\cite{2}.

It is possible to approach the problem of behavior of the entropy from the
other side. We already saw in section~5, that the function of distribution
$f(\om,\, x-\om t)$ for $t\to\infty$ tends in the generalized sense to the
average value~$\ol{f}$, depending only on~$\om$. Let's assume that
$$
  S_\infty =- (2\pi)^n\intl_{\mR^n} \ol{f} \ln\ol{f}\, d^n\om.
  \eqno {(7.2)}
$$
This expression can be interpreted as an entropy in steady equilibrium
state. By the way, formula~(7.2) can be obtained by theorem~\ref{teo2},
using expression $g=\ln f$.

The following inequality
$$
S\le S_\infty,
\eqno {(7.3)}
$$
expresses {\it the law of degradation of energy} of thermodynamics for
irreversible processes.

To prove (7.3) we shall fix the value of $\om$ and then denote
$f(\om,\,x)$ as~$\rho(x)$. Now, we establish the inequality
$$
  \ol {\rho\ln\rho} \ge \ol {\rho} \ln\ol {\rho}.
$$
It is in turn equivalent to the discrete inequality
$$
  \sum \rho_i\ln\rho_i\ge \Bigl (\sum\rho_i\Bigr) \ln\sum\rho_i
$$
for the positive $\rho_i$, which is the special case of the Jensen
inequality for a convex function $\rho\mapsto\rho\ln\rho$.\qed

As Poincar\'{e} has noted~\cite{12}, the values of entropy can be compared
only in the states of steady equilibrium.

Let's consider a simple, but instructive example. Let vessel~$\Pi$ be
divided by a barrier into two parts and the gas be initially concentrated
in one of the parts of~$\Pi$, and in heat balance. Its entropy we shall
denote by~$S_-$. Now we remove the barrier. The gas will extend
adiabatically, uniformly filling (by theorem~\ref{teo1}) the whole volume
of~$\Pi$. Let~$S_+$ be the entropy of gas after the establishment of heat
balance, that will happen after an infinite time. Accordingly to~(7.3),
${S_-\le S_+}$. Moreover, it is possible to show that this case involves
the following simple relation:
$$
  S_+ = S_- +\ln\frac {v_+} {v_-},
  \eqno {(7.4)}
$$
where $v_-(v_+)$ is the volume of $\Pi_-$($\Pi$).

Indeed, let $f_-$ be a density of distribution of gas, being in heat
balance in vessel~$\Pi_-$; this function depends only on velocity~$\om$.
Then, obviously, $S_- =-v_-J $, where
$$
  J = \intl_{\mR^n} f_-\ln f_- \, d^n\om.
$$
After removal of the barrier the equilibrium is broken and the density
$f(\om,\,z)$ now depends on the point $z\in\Pi$: in the initial moment
$f=f_-$, if ${z\in\Pi_-}$ and~$f=0$, if $z\in\Pi\setminus \Pi_-$.
Theorem~\ref{teo2} implies that for ${t\to\infty}$ density~$f$ tends in
generalized sense to the average value
$$
  \ol {f} = \frac {1} {v_+}\intl_{\Pi} f\,d^nz=\frac{v_-}{v _ +} f_-.
$$
Finally,
$$
  \alig {
  S _ + & = -v_+ \intl_{\mR^n} \frac{v_-}{v_+}f_-\ln\Bigl(\frac{v_-}{v_+}f_-\Bigr)
  d^n\om = \\
  {}&= -v_- J-v_-\intl_{\mR^n}f_-d^n\om\,\ln\frac{v_-}{v_+}
  = S_- +\ln\frac {v_+}{v_-}.
 }
$$

The formula (7.4) coincides with the well-known formula of the increase of
entropy for the process of free expansion of gas into the void. However,
we have obtained this formula, not basing on the laws of thermodynamics,
but using only the law of motion of ideal gas as a collisionless medium.

\section{The change of the vessel shape}

Will our deductions change if we replace the rectangular parallelepiped
with a surface of arbitrary shape? This problem has the principal value,
not only from the point of view of thermodynamics.

The matter is that trajectories of particles of gas are essentially the
light rays. Therefore, the problem can be reformulated in terms of
geometrical optics. Let's put a light source (probably distributed) inside
the closed reflecting surface. The question is, will the illumination
inside this surface be constant or will it depend on the coordinates?
Actually the related problem also arises if we consider the radiation in a
closed volume with the beam approach (see, for example, \cite{21}). The
above question is traditionally answered positively. But, in this
situation together with the arguments of dynamical character the
requirement of heat balance is usually used. By the way, Poincar\'{e}
himself made contradictory statements on this subject (see~\cite{22}).

Actually the answer is certainly negative, and it is easy to understand,
bearing in mind the presence of focal points and caustics. As a simple
example, it is possible to consider the vessel having a shape of ellipse
and put the light source in one of the focuses. It is easy to understand,
that the resulting illumination will concentrate on the transversal line
of ellipse (by the way, it is unstable). If the light source is not
located in the focus, the illumination intensity inside the ellipse will
be variable. The existence of a limit distribution in this case (as well
as in any other integrable problem) follows from theorem~\ref{teo2}. If
billiard is not integrable, the existence of a limit density of
distribution presents an interesting problem.

In some sense, any billiard can be arbitrarily precisely approximated by
an integrable billiard. On a plane it is a polygon, with angles comparable
to~$\pi$ (see, for example,~\cite{23}). Except for the integral of
energy, such systems allow the integral in the form of velocity
polynomial. For example, in case of the rectangle the additional integral
has the power~2 (squared projection of velocities on any of the sides is
preserved). It is easy to understand, that the billiards showed in Fig.~3
share the same property. Using the integrability of these systems, with
the help of theorems~\ref{teo1} and~\ref{teo2}, it is possible to study
diffusion of ideal gas as a collisionless medium in vessels of the
indicated shape and, in particular, to study the processes of diffusion
and mixture of gases in vessels with barriers.

\fig<bb=0 0 100.3mm 30.8mm>{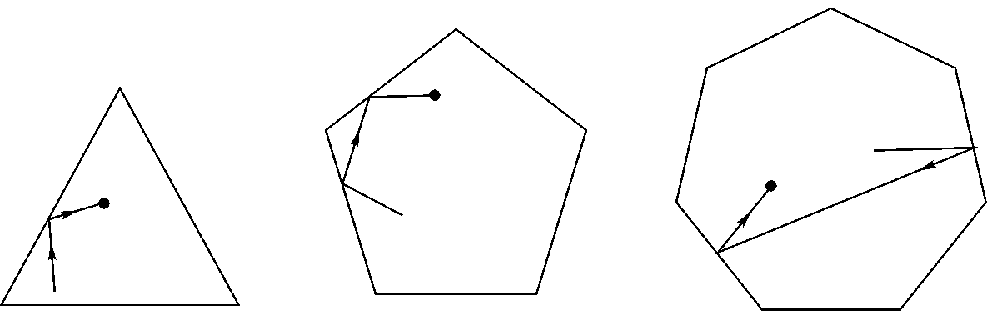}

So, the indicated mechanism of irreversible diffusion of collisionless
medium is not universal. However, for the rarefied gases it should be
considered together with the mechanism of chaotization based on molecular
collisions. According to Poincar\'{e}, after a time, sufficient enough to
let every particle go the whole length of the vessel several times, but
short enough for the collisions not to be too numerous, the mode will be
established in gas that corresponds to the equilibrium state of a
collisionless continuous medium. But this equilibrium will not be final,
the collisions will try to break it, and only after a much larger time
interval the gas will at last reach the final heat balance.

By the way, ``a strange kinetics'' with participation of the dynamic demon
of Maxwell discovered in papers~\cite{13,14}, is also based on
consideration of a collisionless medium. The numerical calculations show
that the density of gas is not equalized in vessels, made up from
scattering billiards. As we just have shown, the similar effect occurs for
integrable billiards. The strangeness of kinetics disappear as soon as we
begin to take into account the interaction of particles of gas.

In conclusion of the paper we shall make two remarks.

Let's consider a nondegenerate and quite integrable Hamiltonian system,
which is described by equations~(2.2) in each invariant domain $D\times
\mT^n$ of the phase space, where $D$ is a domain in $ \mR^n = \{\om \} $.
Let's perturb slightly the Hamiltonian function. Though the perturbed
system is not integrable any more, the majority (in the sense of Lebesgue
measure) of invariant tori do not disappear, but become only slightly
deformed ({\it the Kolmogorov's theorem} on persistence of conditionally\1
periodic motions). When the Hamiltonian function is sufficiently smooth
the dynamics on the invariant set of large measure will be described again
by equations~(2.2), but now $\om\in D\setminus M $, where $\mes M\to 0$,
when the perturbation disappears~\cite{15}. In this situation
theorems~\ref{teo1} and~\ref{teo2} will be still valid, but the density of
distribution of probabilities~$f$ must be equal to zero on the direct
product $M\times \mT^n$ (in the gap between {\it Kolmogorov tori}). Thus,
it is possible to speak about the diffusion of perturbed Hamiltonian
system on the invariant Kolmogorov set. Note that $D\setminus M$ has the
structure of a Cantor set; in particular, it is nowhere dense in~$D$.

Note also that the limit density of distribution~$\ol{f}$ can be obtained
by averaging of function~$f$ over trajectories of system~(2.2). Let's
assume that
$$
  \wt f(\om,\, x) = \lim_{T\to\infty}\frac 1T \intl_0^T f(\om,\, x-\om t)\, dt.
  \eqno {(8.1)}
$$
The Weyl theorem implies that this limit exists for all phases~$x$ and
$\wt f(\om,\,x)=\ol{f}(\om)$ for all {\it nonresonance} sets of
frequencies $\om=(\om_1\dts \om_n)$. We know that the frequencies
$\om_1\dts \om_n$ are {\it in resonance}, if $k_1\om_2 +\ldots+k_n\om_n=0$
for certain integer~$k_s$, not all of which are equal to zero. Since the
resonance sets $\om\in\mR^n$ amount to a set of zero Lebesgue measure,
then, from the point of view of the measure theory, functions~$\wt f$
and~$\ol{f}$ are equivalent. The function~$\wt f$ is continuous for
nonresonance values~$\om$ and, in general, is discontinuous on a set of
the resonance tori~\cite{24} (as a classical example of Riemann function,
continuous in irrational and discontinuous in rational points of the real
axis).

Let's consider now a dynamical system, more general than~(2.2). It is
defined by the autonomous system of differential equations
$$
  \dot x=v (x, \, \om), \qq \dot\om=0
  \eqno {(8.2)}
$$
on a direct product $\mP =\Lm\times \mR^m$, where $\Lm = \{x \}$ is a
compact $n$\1dimentional manifold; $v$ is a smooth vector field on~$\Lm$
with an invariant measure $d\mu =\lm (x, \, \om) \, d^n x $:
$$
  \sum\pt {(v_i\lm)} {x_i} =0.
$$
In contrast to (2.2), the dimensions $m$ and $n$ do not coincide. For
example, for ${m=1}$ equations~(8.2) describe the Hamiltonian systems with
compact energy manifolds; the total energy of the system acts as
variable~$\om$.

Let $f(\om,\, x)$ and $g(\om, \, x)$ be the functions, summable with a
quadrate, defined on the phase space~$\mP$. Following the construction
of~(3.2), we consider the function
$$
K(t) = \intl_\mP f(\om,\,g^{-t}(x,\,\om)g(\om,\,x))\,d^m\om\,d\mu,
\eqno {(8.3)}
$$
where $g^t$ is a phase flow of system (8.2). For the purposes of
justification of the kinetics of a collisionless continuous medium in the
general case, we shall prove the following statement: if for almost
all~$\om$ the dynamical system $\dot x=v (x, \, \om)$ is ergodic on~$\Lm$,
then
$$
  \lim_{t\to\infty} K(t) = \intl_{\mR^m} \ol{f} \ol{g}\, \mes \Lm\, d^m\om,
  \eqno {(8.4)}
$$
where
$$
  \ol {(\cdot)} = \frac {1} {\mes\Lm} \intl_{\Lm} (\cdot)\, d\mu, \q
  \mes\Lm =\intl_{\Lm}\, d\mu.
$$
The relation (8.4) contains the formula (5.1) as a subcase.

Note that if for almost all $\om$ the dynamical system on~$\Lm$ has the
{\it mixing} property the formula~(8.4) is, certainly, valid. Indeed, for
fixed~$\om$ (by definition of mixing)
$$
  \intl_\Lm f(\om,\,g^{-t}(x,\,\om))g(\om,\,x)\,d\mu\to \frac {1} {\mes\Lm}
\intl_\Lm f \, d\mu\intl_\Lm g \, d\mu
  \eqno {(8.5)}
$$
for $t\to\infty $ (see, for example, \cite{25}).
Then we should integrate this relation over~$\mR^m$.

It is important to emphasize, that the system $\dot x =\om $ on torus is
not mixing. In this case when proving of~(8.4), we essentially use an
averaging on~$\om$ and the property of Fourier transform.

Note that for ergodic systems without mixing relation~(8.5) is valid only
for {\it Cesaro convergence}, when the left-hand part is time averaged (as
in~(8.1)) (see~\cite{25}). However, the time averaging (characteristic for
the ergodic theorems by von Neumann and Birkhoff) should be replaced with
the averaging by~$\om$ in our case. Note, that in Kac's circular model
there is also an additional averaging over all the possible states of the
set~$M$, which is used in description of the dynamics of white and black
balls.

Let $\Lm =\mT^n$ and the field $v$ does not depend on $x$. Then it is
possible to set $\lm=1$ and, consequently, ${\mes\Lm = (2\pi)^n}$.
Formula~(8.4) is obviously valid in this case if we assume that
$m$\1dimensional surface ${\om\mapsto v (\om)}$ in ${\mR^n = \{v \}}$ is
transversal to the resonance planes ${(k, \, v) =0}$,
$k\in\mZ^n\setminus\{0\}$. The problem of validity of~(8.4) in the general
case is still open.

The paper is prepared with the financial support of RFBR (99-01-01096) and
the ``Leading scientific groups'' grant (00-15-96146).

\newpage

\thispagestyle{empty}
\mbox{}

\end{document}

%% file: rcdhyph.tex

\begingroup
\catcode`\+\active\gdef+{\mathchar8235\nobreak\discretionary{}%
 {\usefont{OT1}{cmr}{m}{n}\char43}{}}
\catcode`\-\active\gdef-{\mathchar8704\nobreak\discretionary{}%
 {\usefont{OMS}{cmsy}{m}{n}\char0}{}}
\catcode`\=\active\gdef={\mathchar12349\nobreak\discretionary{}%
 {\usefont{OT1}{cmr}{m}{n}\char61}{}}
\endgroup
\mathcode`\+=32768
\mathcode`\==32768 
\mathcode`\-=32768


\makeatletter
\newcommand{\1}{\protect\nobreakdash-\hspace{\z@}}
\newcommand{\2}{\protect\nobreakdash--\hspace{\z@}}
\renewcommand{\f}{\kern1.5\p@\protect\nobreakdash--\kern1.5\p@\hspace{\z@}}
\makeatother


\let\ra=\longrightarrow         \let\la=\longleftarrow
\let\Ra=\Rightarrow             \let\La=\Leftarrow
\let\Lr=\Leftrightarrow         
\let\Lra=\Longrightarrow        \let\Lla=\Longleftarrow
\let\lra=\leftrightarrow        \let\Llra=\Longleftrightarrow
\let\llra=\longleftrightarrow   \let\ora=\overrightarrow 
\let\ua=\uparrow                \let\da=\downarrow
\let\xr=\xrightarrow            \let\xl=\xleftarrow 

\let\ud=\d               \let\d=\partial           \let\ob=\overbrace
\let\ub=\underbrace      \let\ol=\overline         \let\ul=\underline
\let\ds=\displaystyle    \let\ts=\textstyle        \let\scs=\scriptstyle
\let\ctl=\centerline     \let\nn=\noindent         
\let\t=\text
\let\q=\quad             \let\qq=\qquad            \let\lt=\limits
\let\bs=\boldsymbol      \let\x=\times
\let\de=\doteq           \let\wt=\widetilde        \let\sps=\supset
\let\wh=\widehat         \let\sbs=\subset          \let\sbq=\subseteq
\let\fy=\infty           \let\vnth=\varnothing     \let\spq=\supseteq

\let\al=\alpha           \let\bt=\beta             \let\gam=\gamma
\let\dl=\delta           \let\vth=\vartheta        \let\vk=\varkappa
\let\lm=\lambda          \let\sg=\sigma            \let\vfi=\varphi
\let\om=\omega           \let\ta=\theta            \let\vr=\varrho
\let\eps=\varepsilon     \let\zt=\zeta             \let\vsg=\varsigma
\let\nb=\nabla

\let\Gam=\Gamma           \let\Dl=\Delta            \let\Tt=\Theta
\let\Lm=\Lambda          \let\Sg=\Sigma            \let\Om=\Omega
\let\vFi=\varPhi         \let\vTt=\varTheta        \let\vSg=\varSigma
\let\vOm=\varOmega       \let\vP=\varPsi           \let\vLm=\varLambda
\let\vPi=\varPi          \let\vUp=\varUpsilon      \let\vGm=\varGamma
\let\vDl=\varDelta       \let\vXi=\varXi           \let\Gm=\Gamma
\let\Up=\Upsilon         \let\Fi=\Phi